\documentclass{aa}

\usepackage[varg]{txfonts}
\usepackage{graphicx}

\usepackage{hyperref}
\usepackage{epstopdf}

\bibpunct{(}{)}{;}{a}{}{,} 

\usepackage{mathtools}

\usepackage{siunitx}

\usepackage[version=3]{mhchem}

\newcommand{\Msun}{\ensuremath{M_{\odot}}}
\newcommand{\about}{\ensuremath{\mathord{\sim}}}	
\newcommand{\ssup}[1]{\ensuremath{^{\mathrm{#1}}}}  
\newcommand{\diff}{\ensuremath{{\mathrm{d}}}}		

\begingroup
  \catcode`\_=\active
  \gdef_#1{\ensuremath{\sb{\mathrm{#1}}}}	
\endgroup
\mathcode`\_=\string"8000
\catcode`\_=12

\begin{document}

	\title{The influence of magnetic fields, turbulence, and UV radiation on the formation of supermassive black holes}
	
	\author{C. Van Borm\inst{1,2}\thanks{Currently at the Institut f\"{u}r Astrophysik, Georg-August-Universit\"{a}t G\"{o}ttingen, Friedrich-Hund-Platz 1, 37077 G\"{o}ttingen, Germany}
	\and M. Spaans\inst{1,3}}

	\institute{Kapteyn Astronomical Institute, University of Groningen, PO Box 800, 9700 AV, Groningen, The Netherlands
		\and \email{borm@astro.rug.nl}
		\and \email{spaans@astro.rug.nl}}

	\date{Received <date> / Accepted <date>}

	\abstract
  	{The seeds of the supermassive black holes (SMBHs) with masses of \about \SI{e9}{\Msun} observed already at $z \sim 6$ may have formed through the direct collapse of primordial gas in $T_{vir} \gtrsim \SI{e4}{K}$ halos, whereby the gas must stay hot (\about \SI{e4}{K}) in order to avoid fragmentation.}
  	{The interplay between magnetic fields, turbulence, and a UV radiation background during the gravitational collapse of primordial gas in a halo is explored; in particular, the possibilities for avoiding fragmentation are examined.}
  	{Using an analytical one-zone model, the evolution of a cloud of primordial gas is followed from its initial cosmic expansion through turnaround, virialization, and collapse up to a density of \SI{e7}{cm^{-3}}.}
  	{It was found that in halos with no significant turbulence, the critical UV background intensity ($J_{21}\ssup{crit}$) for keeping the gas hot is lower by a factor \about 10 for an initial comoving magnetic field $B_0 \sim \SI{2}{nG}$ than for the zero-field case, and even lower for stronger fields. 
In turbulent halos, $J_{21}\ssup{crit}$ is found to be a factor \about 10 lower than for the zero-field-zero-turbulence case, and the stronger the turbulence (more massive halo and/or stronger turbulent heating), the lower $J_{21}\ssup{crit}$.}
   	{The reduction in $J_{21}\ssup{crit}$ is particularly important, since it exponentially increases the number of halos exposed to a supercritical radiation background.}

               
	\titlerunning{The formation of supermassive black holes}
	\maketitle

\section{Introduction} \label{sec:introduction}

Several very bright quasars have been detected already at $z > 6$. This suggests that some supermassive black holes (SMBHs) with masses of $\simeq \SI{e9}{\Msun}$ already existed when the Universe was less than \SI{1}{Gyr} old \citep{fan2006}. Explaining how such massive SMBHs could have assembled so soon after the Big Bang presents quite a challenge. The main questions concern how and when the seeds of these SMBHs formed and how their subsequent growth proceeded. 

Several pathways leading to the formation of seed black holes (SBHs) have already been proposed. 
One group of scenarios suggests that SBHs formed via the direct collapse of metal-free/very metal-poor gas in halos with $T_{vir} \gtrsim \SI{e4}{K}$, at redshifts \about $5-10$, resulting in SBHs with $M \sim \num{e4}-\SI{e5}{\Msun}$ \citep[see, e.g.,][]{bromm2003,koushiappas2004,begelman2006,lodato2006,spaans2006,schleicher2010b,latif2013}. For efficient gas collapse to occur, fragmentation must be suppressed, which is possible if the gas in the halo is kept hot (large Jeans mass). Hence, the formation of \ce{H2} must be inhibited so cooling can occur only through atomic \ce{H}, as otherwise \ce{H2} cooling will lower temperatures to \about \SI{200}{K}. 

Several mechanisms have been suggested that suppress \ce{H2} cooling. The most accepted of these mechanisms requires a critical level of Lyman-Werner radiation to photo-dissociate \ce{H2} and keep its abundance very low. The critical intensity needed to suppress \ce{H2} in the massive halos where direct gas collapse can occur is large compared to the expected cosmic UV background at the relevant redshifts. However, the distribution has a long bright-end tail, and halos irradiated by supercritical intensities would be a small subset of all halos \citep{dijkstra2008}. Another mechanism proposes that the dissipation of a sufficiently strong magnetic field can heat the gas in the halo to \about \SI{e4}{K}, which causes \ce{H2} to be destroyed by collisional dissociation \citep{sethi2010}.

A variety of mechanisms exist for generating magnetic fields early in the Universe, both before and after recombination \citep[for a review, see, e.g.,][]{widrow2012}, and also for amplifying an existing field.
In the case of a collapsing halo, the most important ones are gravitational compression, the small-scale turbulent dynamo, the large-scale dynamo in protostellar and galactic disks, and the magneto-rotational instability (MRI). 

Gravitational compression increases the magnetic field as $B \propto \rho_{b}^{\alpha}$ when the field is coupled to the gas. For spherically symmetric collapse, $\alpha = 2/3$, but if the collapse proceeds preferentially along one axis, the scaling is closer to $\alpha = 1/2$. In realistic cases, intermediate values are often found \citep[e.g.,][]{schleicher2009,hocuk2012}.

Non-helical turbulent flows can act as small-scale dynamos, which produce disordered, random magnetic fields \citep{kazantsev1968}. The magnetic field amplification results from the random stretching and folding of the field lines by the turbulent random flow.
During gravitational collapse, turbulence is generated by the release of gravitational energy and the infall of accreted gas on the inner, self-gravitating core. This means that, in the context of star and galaxy formation, a strong tangled magnetic field may already be generated during the collapse phase by the small-scale dynamo \citep{schleicher2010}. 
For the formation of SBHs, it implies that the existence of an accretion disk may cause the magnetic field to be further amplified (by a large-scale dynamo and/or the MRI) which provides additional stability and hence reduces fragmentation.

\section{The model} \label{sec:the_model}
The evolution of a cloud of primordial gas is followed from initial expansion to a high-density core, using a one-zone model, in which the physical variables involved are regarded as those at the cloud center.
The model uses standard cosmology, with cosmological parameters given by WMAP7 \citep{larson2011}. 

\subsection{Density evolution}
The spherical collapse model for a top-hat overdensity is used to compute the matter density. At the moment of turnaround, the gas decouples from the dark matter and becomes self-gravitating.
Any effects due to rotation are neglected for simplicity. The baryonic matter collapse is expected to proceed like the Larson-Penston similarity solution \citep[for the isothermal case;][]{larson1969,penston1969}, as generalized to polytropic cases by \citet{yahil1983}. According to this solution, the cloud consists of two parts, a central core region and an envelope. The central core region has a flat density distribution, whereas the density in the envelope decreases outwards. The size of the central flat region is roughly given by the local Jeans length, $\lambda_{J} = c_{s} \sqrt{\pi /\left(G \rho_{m}\right)}$,  with $c_{s} = \sqrt{\gamma k_{B} T / \left(\mu m_{H}\right)}$ the sound speed in the central region and $\rho_{m}$ the total matter density in the central region. 
The collapse in the core proceeds approximately at the free-fall rate, although additional heat input due to magnetic energy dissipation, for example, may delay gravitational collapse. 
The mean baryonic density evolution in the central part is described by $\dot{\rho_{b}} = \rho_{b}/t_{ff}$,
where $t_{ff}$ is the free-fall time $t_{ff} = \sqrt{3\pi / \left(32G\rho_{m}\right)}$.

\subsection{Chemical network}
The species that are included in the chemical network of this model are \ce{H}, \ce{H+}, \ce{H-}, \ce{H2}, \ce{H2+}, and \ce{e-}; \ce{HD} or other molecules involving deuterium are not included.
Reactions with \ce{He} are not taken into account, but \ce{He} is considered in the calculation of the mean molecular mass. The \ce{He} mass fraction is taken to be \about 0.248 (corresponding to an abundance $x_{He} \approx 0.0825$) and stays constant throughout the time integration. The fractional abundances of \ce{H}, \ce{H2}, and \ce{e-} are explicitly followed during the integration.

The evolution of the fractional abundance of electrons $x_{e}$ is given by the equation \citep[see, e.g.,][for further details]{peebles1993}
\begin{align}
\frac{\diff x_{e}}{\diff t} &= \left[\beta_{e} x_{HI} \exp{\left(-\frac{h\nu_{\alpha}}{k_{B} T_{cmb}}\right)} - \alpha_{e} x_{e}^2 n_{H}\right]C + \gamma_{e}(T) x_{HI} x_{e} n_{H}.
\end{align}
The first term represents the recombination and photo-ionization of the primordial plasma, the second term is the collisional recombination term, and the third term represents collisional ionization (\ce{H + e- -> H+ + 2e-}). 

The evolution of the fractional abundance of molecular hydrogen $x_{H_2}$ is given by the equation \citep{shang2010,sethi2010}
\begin{align}
\frac{\diff x_{H_2}}{\diff t} &= k_{m} x_{e} x_{HI} n_{H} - k_{des} x_{H_2} n_{H}, \\
\intertext{where}
k_{m} &= \frac{k_9 k_{10} x_{HI} n_{H}}{\splitfrac{k_{10} x_{HI} n_{H} + k_{\gamma} + \left(k_{13}+k_{21}\right)x_{e} n_{H}}{ + k_{19} x_{e} n_{H} + k_{20} x_{HI} n_{H} + k_{25}}}, \\
k_{des} &= k_{15} x_{HI} + k_{17} x_{p} + k_{18} x_{e} + k_{28} f_{sh} / n_{H}.
\end{align}
Here, $k_{m}$ is the net rate of formation of \ce{H2} through the \ce{H-} channel, $k_{des}$ is the net destruction rate of \ce{H2}, and $k_{\gamma}$ is the destruction rate of \ce{H-} by CMB photons. The reaction rates can be found in Appendix A of \citet{shang2010}, numbered as above. 
For sufficiently large column densities, \ce{H2} can shield itself from radiation in the Lyman-Werner bands. The self-shielding factor $f_{sh}$ is given by \citet{draine1996}.
The updated $f_{sh}$ from \citet{wolcott-green2011} results in decreased shielding and may thus enhance our results.

\subsubsection{Radiation background}
A sufficiently intense UV radiation background can either directly photo-dissociate \ce{H2}, or photo-dissociate the intermediary \ce{H-}. The relevant criterion for suppressing \ce{H2} formation is that the photo-dissociation timescale must be shorter than the formation timescale, which results in $J\ssup{crit} \propto \rho$. 
The intensity is written as $J_{21}$, which denotes the specific intensity just below \SI{13.6}{eV} in the units of \SI{e-21}{erg.cm^{-2}.sr^{-1}.s^{-1}.Hz^{-1}}. 
Here, two different UV spectra are considered. They are both Planck spectra with a blackbody temperature of either $T_* = \num{e4}$ or \SI{e5}{K} (T4 and T5, respectively). The T4 spectrum is meant to approximate the mean spectrum of a normal stellar population (Pop II), whereas the T5 spectrum is closer to the harder spectrum expected to be emitted by the first generation of stars (Pop III) \citep{tumlinson2000,bromm2001,schaerer2002}.

\subsection{Magnetic field evolution}
Gravitational compression and the small-scale dynamo can amplify the magnetic field strength $B$, while ambipolar diffusion will decrease $B$. 
If the flux-freezing condition applies, the magnetic field depends on the density as $B \propto \rho_{b}^{\alpha}$, where $\alpha$ lies in the range $2/3 - 1/2$. Hence, $B$ will increase during gravitational collapse. 
It is assumed that gas is continually falling in, so the turbulence generated by accretion will not decay, but is instead constantly replenished. However, this depends on the ambient gas reservoir and may in reality be more complicated.
It has been shown that the injection scale of such accretion-driven turbulence is close to the size of the system under consideration (thus, $\lambda_{J}$) \citep{klessen2010,federrath2011}.
The turbulent velocity on the injection scale is expected to be comparable to the velocity of the infalling gas, and for a roughly isothermal density profile, the free-fall velocity is independent of radius. So, while the injection scale changes during the collapse, the injected velocity $v_{in}$ stays the same and is approximately equal to the virial velocity \citep{greif2008,wise2007,wise2008}. On scales smaller than the injection scale, the turbulent velocity is expected to scale as $v \propto l^{\beta}$, with $\beta = 1/3$ for Kolmogorov turbulence (incompressible gas) and $\beta = 1/2$ for Burgers turbulence (strongly compressed gas). 

The magnetic field on a scale $l$ typically grows exponentially on the eddy turnover time $t_{ed} = l/v$, where $v$ is the turbulent velocity on the scale $l$. However, the magnetic field will saturate when the magnetic energy corresponds to a fraction $Rm_{cr}^{-1}$ of the kinetic energy; $Rm_{cr}$ is the critical magnetic Reynolds number. If $Rm > Rm_{cr}$, the magnetic field grows through dynamo action. The maximum magnetic field strength is thus given by $B_{max} = \sqrt{4 \pi \rho_{b} v^2 / Rm_{cr}}$ \citep{subramanian1998}.
However, the exact value of the saturation field strength is still somewhat uncertain.
Once $B > B_{max}$ on a certain scale, it is no longer amplified by the small-scale dynamo; however, it is still amplified by gravitational compression ($\propto \rho^{\alpha}$). It can then in principle increase above the saturation level (which only increases $\propto \rho^{1/2}$), but in this case it is subject to turbulent decay. On a given scale, this decay will probably also occur on $t_{ed}$. As a result, the magnetic field strength tends to stay close to $B_{max}$ on that scale.

The most important contribution to the total magnetic energy comes from the integral scale, the scale on which the magnetic field is largest. \citet{schleicher2010} have shown that in an atomic cooling halo, the integral scale increases very rapidly to the maximum scale on which the magnetic field is coherent after the start of the simulation. For this reason, only the evolution of the magnetic field at this scale of maximal coherence is followed here. Since the magnetic field is distorted by the gravitational collapse, the largest possible coherence length is always smaller than the Jeans length by some factor $f_d$; its precise value is uncertain. We adopted a fiducial value of \num{0.1}; changing this by a factor of a few does not significantly affect the results.

Ambipolar diffusion (AD) is important in a mostly neutral medium where a tangled magnetic field is present. The AD heating rate can be estimated as $L_{AD} \approx \eta_{AD} B^2 / \left(4\pi l_B^2\right)$ \citep{shang2002,schleicher2008},
where $\eta_{AD}$ is the AD resistivity \citep{pinto2008a,pinto2008b} and $l_B$ is the coherence length of the magnetic field, which is approximated by the minimum of the Alv\'{e}n damping scale and the integral scale.

The evolution of the magnetic field energy $E_{B} = B^2 / 8\pi$ is then calculated as
\begin{equation}
\frac{\diff E_{B}}{\diff t} =
\begin{cases}
	2 \alpha E_{B} \ln{\dot{\rho_{b}}} - L_{AD} & z \geq z_{vir}, \\
	2 E_{B} \left( \ln{\dot{B}} \right)_{dynamo} & z_{vir} > z.
\end{cases}
\end{equation}

\subsection{Thermal evolution}
The evolution of the temperature $T$ is given by the equation \citep[see, e.g.,][]{peebles1993,sethi2008}
\begin{equation}
\frac{\diff T}{\diff t} = \frac{\gamma - 1}{\rho_{b}} \left[T \frac{d\rho_b}{dt} + \frac{\mu m_{H}}{k_{B}} \left(L_{h} - L_{c}\right) \right] + k_{C} x_{e} \left(T_{cmb} - T\right). 
\end{equation}
The first term is the adiabatic heating/cooling rate due to collapse/expansion, the second term incorporates various other heating (magnetic energy dissipation by AD and turbulent dissipation, discussed below) and cooling (through \ce{H2} and \ce{H}) volume rates, and the third term represents Compton heating/cooling.

Part of the turbulent energy will go to driving the small-scale dynamo, and part of it will be transferred from large scales to smaller ones in a cascade process, until it is dissipated by viscosity at small enough scales. 
The rate (per unit mass) at which energy is injected into the system is $\epsilon_{in} = E_{in}/m / t_{ed}(\lambda_{J}) = v_{in}^3 / \left(2 \lambda_{J}\right) $\citep{shu1992}.
The volume heating rate from the accretion-driven turbulence (ADT) can then be estimated as $L_{ADT} = f_t \rho_{b} \epsilon_{in}$, where $f_t$ is the fraction of the injected energy that is dissipated. Its fiducial value is taken to be \num{0.1}; changing this by a factor of a few does not significantly affect the results.

\paragraph{}
The model is initialized at a redshift of 800. The dissipation of magnetic fields into the IGM after recombination can significantly influence its temperature and ionization, so this early start provides the proper initial conditions.
We follow the evolution of a \SI{e9}{\Msun} halo that virializes at $z=10$. 
The radiation background is switched on at turnaround ($z \approx 15$); an earlier or slightly later turn-on is also possible, but does not change the results significantly.
The integration is stopped when $n_b \approx \SI{e7}{cm^{-3}}$. The model is not suitable for higher densities; this would require the inclusion of additional physical processes, e.g., three-body interactions which increase the \ce{H2} formation rate.

\section{Results} \label{sec:results}
\begin{figure}
	\resizebox{\hsize}{!}{\includegraphics{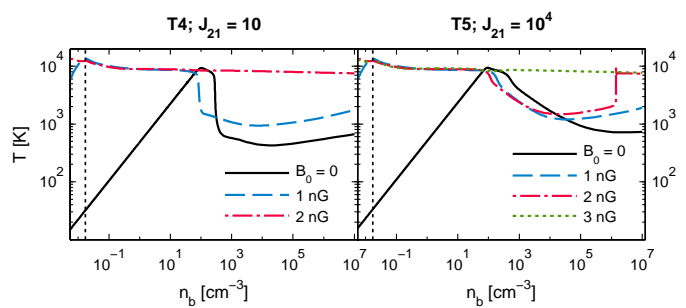}}
	\caption{Gas temperature as a function of density for a T4 background with $J_{21} = 10$ (left) and a T5 background with $J_{21} = \num{e4}$ (right), for several different initial magnetic field strengths (no turbulence). The dotted vertical line indicates virialization.}
	\label{fig:NoTurb_T4T5_B}
\end{figure}

After the radiation background is turned on, \ce{H2} is destroyed rapidly. It cannot self-shield as the density is too low; this happens even for a low intensity of $J_{21} = 1$. The cooling is dominated by \ce{H}, so the temperature is high and the ionized fraction becomes elevated, which in turn aids the formation of \ce{H2}.
For intensities that are low enough, \ce{H2} succeeds in reforming, and becomes the dominant coolant. The density at which this occurs depends on the radiation intensity, and is higher for higher intensities. However, at a certain intensity \ce{H2} cannot reform fast enough to become an important coolant. 
%
If turbulence is unimportant and the initial magnetic field is $\leq\! \SI{0.01}{nG}$ (virtually identical to the zero-field case), this critical intensity is found to be $10 < J_{21} \leq \num{e2}$ for a T4 background and $\num{e4} < J_{21} \leq \num{e5}$ for a T5 background. Similar results are found for $B_0 = \SI{1}{nG}$, always assuming $\alpha = 2/3$.
However, if the magnetic field is increased to \SI{2}{nG}, \ce{H2} never becomes an important coolant for a T4 background with $J_{21} = 10$, while for $J_{21} = 1$ an instability occurs at high densities. Here, the AD heating becomes too strong to be compensated by the \ce{H2} cooling and the temperature suddenly increases, because much of the \ce{H2} is destroyed by collisional dissociation and \ce{H} cooling is still very strongly suppressed below \about \SI{8e3}{K}. The gas stays hot afterwards, with cooling dominated by \ce{H}. 
For a T5 background with the same $B_0$, a similar instability occurs for $J_{21} = \num{e4}$, while with $ B_0 = \SI{3}{nG}$ the \ce{H2} fraction is never large enough for significant cooling at this intensity. 
Thus, when turbulence is not important, a larger initial magnetic field decreases the critical intensity required to keep the gas in the halo hot, as illustrated in Fig.~\ref{fig:NoTurb_T4T5_B}.

During collapse, turbulence quickly brings the magnetic field strength to $B_{max}$ through the small-scale dynamo, either by amplifying smaller fields or by draining energy from larger fields. Since $B_{max}$ only increases as $\propto n_b^{1/2}$, heating from turbulent dissipation always dominates over AD heating, and because \ce{H2} cooling grows more steeply with density than turbulent heating does, the gas always cools through \ce{H2} when there is no radiation background present. Since heating is dominated by turbulence, halos with different $B_0$ converge to approximately the same evolutionary track; the turbulence has a moderating effect.

\begin{figure}
	\resizebox{\hsize}{!}{\includegraphics{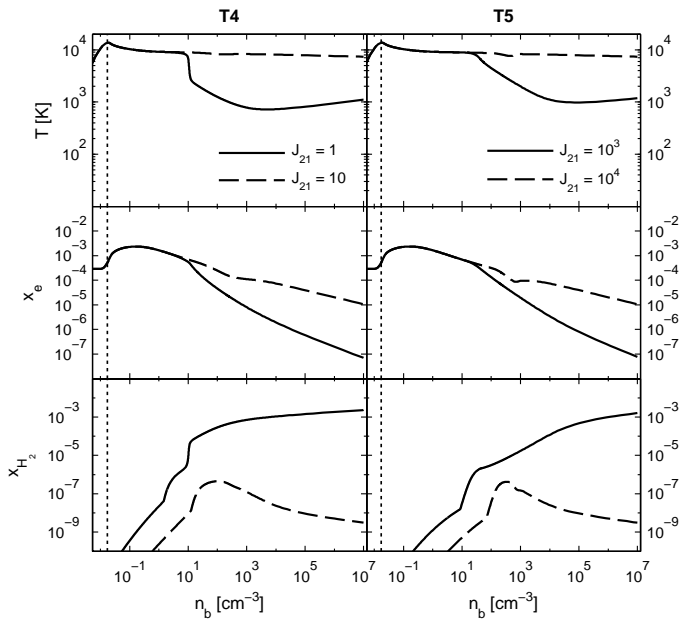}}
	\caption{Gas temperature, electron fraction, and \ce{H2} fraction as a function of density for a turbulent halo with $B_0=\SI{1}{nG}$, a T4 (left) and a T5 (right) background, and different $J_{21}$ as indicated in the top panels. The dotted vertical line indicates virialization.}
	\label{fig:T4T5}
\end{figure}

The effects of different radiation intensities of a T4 and T5 background on the temperature, electron fraction, and \ce{H2} fraction are shown in Fig.~\ref{fig:T4T5} for a turbulent halo with $B_0=\SI{1}{nG}$. However, the critical intensity does not depend on $B_0$ when turbulent effects are important. The critical intensity is found to be $1 < J_{21} \leq 10$ for a T4 background and $\num{e3} < J_{21} \leq \num{e4}$ for a T5 background.

\section{Discussion} \label{sec:discussion}
The effects of magnetic fields and turbulence on the critical intensity of a UV radiation background were examined. Note that these results only hold in a zero or very low metallicity environment, because metals and dust are much more efficient coolants than \ce{H2}. 
For a halo not significantly influenced by turbulence or magnetic fields, the critical intensity was found to be $\num{10} < J_{21}\ssup{crit} \leq \num{e2}$ for a T4 background, and $\num{e4} < J_{21}\ssup{crit} \leq \num{e5}$ for a T5 background. These limits are consistent with those found by \citet{shang2010} and lower by a factor \about 10 than previous estimates by e.g., \citet{omukai2001} and \citet{bromm2003}, most likely as a result of the different \ce{H2} dissociation rates used. 
For $B_0 = \SI{1}{nG}$, these limits do not change; however, when the field is increased to \about \SI{2}{nG}, the critical intensity is lowered by a factor \about 10, and the stronger the field, the lower $J_{21}\ssup{crit}$. 
Such magnetic fields alone do not give rise to sufficient AD heating to overcome \ce{H2} cooling, but in combination with a radiation background they do influence the ability of the gas to stay hot. Note that the amount of AD heating depends on the scaling of $B$ with $\rho$; it is therefore important to obtain a correct model for this relationship.
The current upper limit on the primordial magnetic field is \about \SI{1}{nG} comoving \citep{schleicher2011,trivedi2012}, so a \SI{2}{nG} field would be reached by the \about $2\sigma$ upward fluctuations. 

In a \SI{e9}{\Msun} turbulent halo, the critical intensity was found to be $\num{1} < J_{21}\ssup{crit} \leq \num{10}$ for a T4 background, and $\num{e3} < J_{21}\ssup{crit} \leq \num{e4}$ for a T5 background; these are a factor \about 10 lower than for a halo not affected by turbulence or magnetic fields. Since this is due to the turbulent heating in such halos, larger halos and/or halos with stronger turbulent heating will have an even lower $J_{21}\ssup{crit}$. 
Note that the results for the non-turbulent halos are independent of halo mass. Interestingly, in turbulent halos with $M \gtrsim \SI{e11}{\Msun}$ (depending on the strength of the turbulence), the turbulent heating alone is able to keep the gas hot without any UV background.

The fact that the values of $J_{21}\ssup{crit}$ that have been found here are smaller than previous estimates is quite important. The mean cosmic UV background is expected to be around $J_{21}\ssup{bg} \sim 40$, which is smaller than the critical intensities, especially in the case of a T5 background. However, the background will fluctuate spatially, and thus a fraction of all halos will be irradiated by a supercritical intensity. Then, the lower the critical intensity, the larger the fraction of halos which are suitable candidates for direct SMBH formation; e.g., according to the distribution proposed by \citet{dijkstra2008}, a decrease in $J_{21}\ssup{crit}$ from \num{e4} to \num{e3} means an increase in the fraction of irradiated $T_{vir} \approx \SI{e4}{K}$ halos from negligibly small ($\lesssim \num{e-8}$) to \about \num{e-6}. For $J_{21}\ssup{crit} \about \num{e2}$, the halo fraction even increases to \about \num{e-3}. 
With a sufficiently low $J_{21}\ssup{crit}$, one could argue that this mechanism provided many, if not all, seeds for the SMBHs observed in galaxies today. 

\begin{acknowledgements}
C. Van Borm acknowledges research funding by the German Science Foundation (DFG) under grant SFB 963/1 (project A12).
\end{acknowledgements}

\vspace{-0.4cm}

\bibliographystyle{aa}
\bibliography{SMBHFormation_refs}

\end{document}